\documentclass{article}
\usepackage{enumitem}
\usepackage{spconf,amsmath,graphicx}
\usepackage{multirow}
\usepackage{xcolor}
\usepackage{subcaption}
\usepackage[version=4]{mhchem}
\usepackage{amsfonts}
\usepackage{booktabs}
\usepackage{cite}

\usepackage{mathtools}
\usepackage{physics}
\usepackage{algorithm}
\usepackage{algorithmic}
\usepackage{tikz}
\usetikzlibrary{arrows.meta,positioning,fit,calc,shapes.geometric}
\usepackage{pgfplots}
\pgfplotsset{compat=1.18}
\usepackage[bookmarks=true,breaklinks=true,colorlinks,linkcolor=blue,citecolor=blue,urlcolor=blue]{hyperref}
\usepackage{comment}
\usepackage{xspace}

\newcommand{\design}{{DiffQ}\xspace}

\title{
\textnormal{\design: Unified Parameter Initialization for Variational Quantum Algorithms via Diffusion Models
\vspace{-0.1in}}
}

\name{Chi Zhang$^*$ \quad Mengxin Zheng$^\dagger$ \quad Qian Lou$^\dagger$ \quad Fan Chen$^*$\vspace{-12pt}}
\address{$^\dagger$University of Central Florida, FL, USA\\
$^*$Indiana University, Bloomington, IN, USA\vspace{-12pt}}

\renewenvironment{thebibliography}[1]{%
   \begin{oldthebibliography}{#1}%
     \setlength{\itemsep}{-.4ex}%
}%
{%
   \end{oldthebibliography}%
}

\begin{document}
\maketitle

\begin{abstract}
Variational Quantum Algorithms (VQAs) are widely used in the noisy intermediate-scale quantum (NISQ) era, but their trainability and performance depend critically on initialization parameters that shape the optimization landscape. 
Existing machine learning–based initializers achieve state-of-the-art results yet remain constrained to single-task domains and small datasets of only hundreds of samples. 
We address these limitations by reformulating VQA parameter initialization as a generative modeling problem and introducing \design, a parameter initializer based on the Denoising Diffusion Probabilistic Model (DDPM).
To support robust training and evaluation, we construct a dataset of 15,085 instances spanning three domains and five representative tasks. 
Experiments demonstrate that \design~surpasses baselines, reducing initial loss by up to 8.95 and convergence steps by up to 23.4\%.

\end{abstract}


\vspace{-0.12in}
\section{Introduction}
\vspace{-0.1in}

\textbf{Motivation}.
Variational Quantum Algorithms (VQAs)~\cite{cerezo2021variational} have emerged as leading methods for the noisy intermediate-scale quantum (NISQ) era~\cite{bharti2022noisy}. By combining limited quantum resources with classical optimizers, they reduce reliance on fault-tolerant devices while offering resilience to noise~\cite{cerezo2021variational}, low circuit complexity~\cite{zhang2022variational}, and design flexibility~\cite{kandala2017hardware}. VQAs have already demonstrated success in quantum physics, chemistry, and materials science~\cite{bauer2020quantum, li2019variational, tilly2022variational}. Despite this promise, their scalability remains a central challenge: as system size increases, optimization landscapes flatten exponentially~\cite{mcclean2018barren}, leading to vanishing gradients and poor convergence. Parameter initialization has therefore become a critical strategy~\cite{larocca2024review}, reshaping the landscape to enhance trainability and mitigate suboptimal convergence. Recent deep learning–based initialization methods~\cite{mesman2024nn, liangGraphLearningParameter2024, mengParameterGenerationQuantum2024, leeQMAMLQuantumModelAgnostic2025} define the state of the art, yet they remain task-specific, depend on limited datasets, and are typically validated in narrow settings, constraining their generalizability across diverse VQA applications.

\textbf{Contributions}.  
Inspired by recent advances in Denoising Diffusion Probabilistic Models (DDPMs)~\cite{sohl2015deep}, a state-of-the-art class of generative machine learning methods~\cite{rombach2022high,hoVideoDiffusionModels2022,kongDiffWaveVersatileDiffusion2021,dhariwalDiffusionModelsBeat2021}, we reformulate VQA parameter initialization as a generative modeling task. 
Specifically, we encode optimized VQA parameters into input tensors and train a DDPM to denoise corrupted versions of these tensors, conditioned on textual prompts describing the VQA task. 
Once trained, the model can transform random noise into high-quality parameter tensors guided by task descriptions, which are then used to initialize VQAs. 
This formulation leverages the strong generalization and flexibility of DDPMs to produce robust and transferable initializations. 
We further evaluate \design~on large-scale datasets, demonstrating improved performance and scalability compared with existing methods. 
The key contributions of this work are as follows:

\begin{itemize}[leftmargin=*, topsep=0pt, partopsep=0pt, itemsep=-2pt]
    \item \textbf{Model Architecture}.  
    We introduce \design, a framework for VQA parameter initialization that integrates task- and guidance-specific encodings with a DDPM-based architecture to enhance trainability and enable cross-task generalization. We further outline its training and inference procedures, offering a systematic methodology for deployment across diverse VQA applications.

    \item \textbf{Dataset Construction}.  
    We construct a VQA dataset of 15,085 instances spanning three domains and five tasks, surpassing prior works~\cite{mesman2024nn,liangGraphLearningParameter2024,mengParameterGenerationQuantum2024,leeQMAMLQuantumModelAgnostic2025} that are limited to a single task and only hundreds of samples. The dataset serves as both the benchmark for this study and a public resource to advance future research on VQA parameter initialization.

    \textbf{Performance Evaluation}.  
    We evaluate \design~on the five constructed datasets, demonstrating substantial gains over existing methods, including reductions in initial loss of up to 8.95 and convergence steps by up to 23.4\%.
\end{itemize}
% This paper introduces \design, a novel diffusion-based initializer for VQAs that addresses the limitations of existing methods. Our approach leverages the Denoising Diffusion Probabilistic Model (DDPM) with a 2-level U-Net backbone, OpenCLIP for text-based conditional encoding, and classifier-free guidance. This combination allows us to generate high-quality initial parameters that are universally applicable across various VQA tasks. We conduct extensive experiments on six representative applications: quantum pulse synthesis, the Heisenberg XYZ model, the Transverse-Ising model, the Fermi-Hubbard model, and the hydrogen molecule. Our results demonstrate the effectiveness of \design in reducing initial loss and convergence steps across these diverse applications, highlighting its potential as a universal initializer for VQAs. The key contributions of this work are as follows:

%%%%%%%%%%%%%%%%%%%%%%%
\begin{figure*}
   \centering
   \includegraphics[width=1\linewidth]{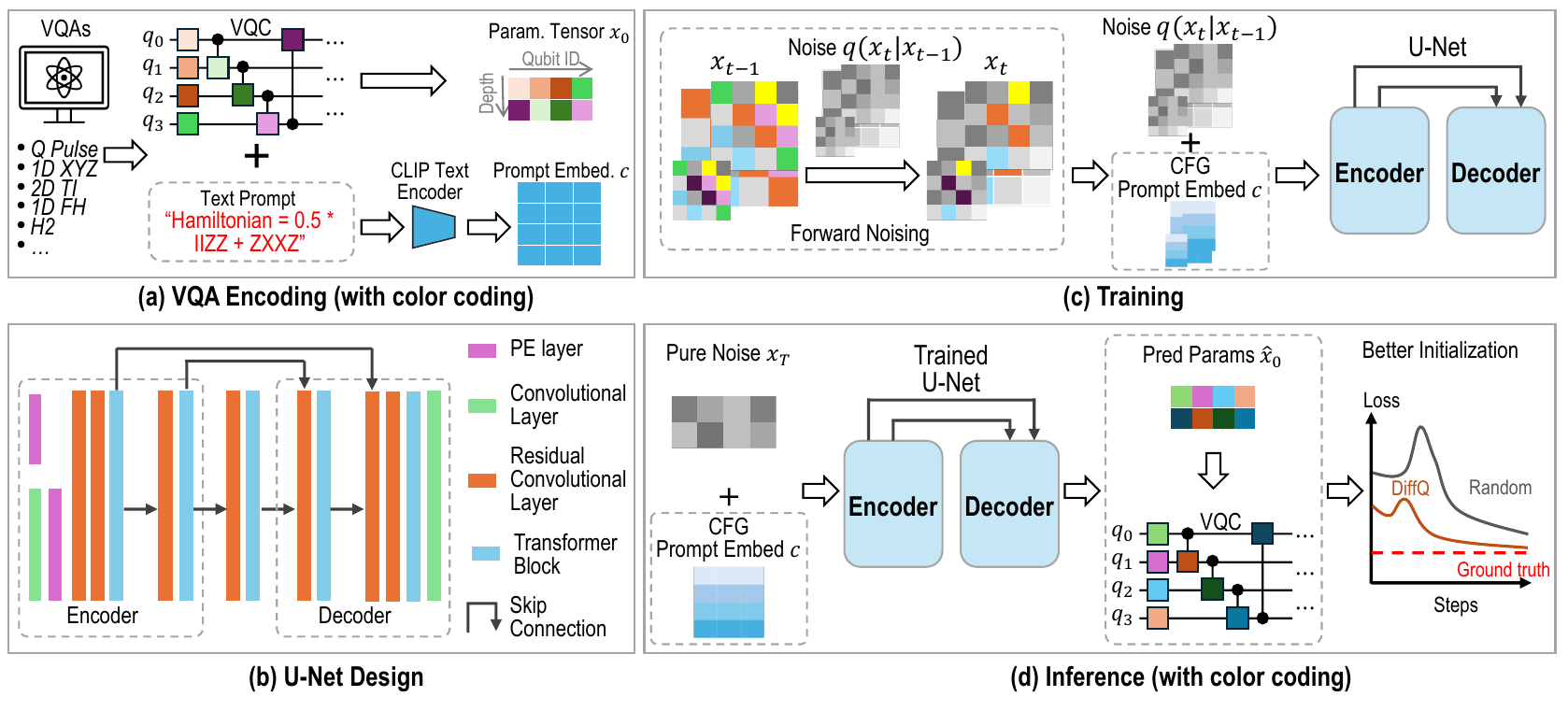}
    \vspace{-0.3in}
    \caption{The \design~framework: (a) encoding of VQA circuit parameters and text-based prompt guidance, (b) U-Net architecture for noise prediction, (c) training procedure of \design, and (d) inference process of \design.}
    \label{fig:design}
    \vspace{-0.26in}
\end{figure*}
%%%%%%%%%%%%%%%%%%%%%%%

\vspace{-0.12in}
\section{\design~Framework}
\label{sec:design}
\vspace{-0.1in}

In this section, we introduce the proposed~\design~framework. We first describe the encoding of VQA circuit parameters and text-based prompt guidance into input tensors. We then present the DDPM-based model architecture for parameter initialization. Finally, we outline the training strategy and inference procedure of~\design.

\vspace{-0.12in}
\subsection{VQA Circuit Parameter and Conditioning Encoding}
\vspace{-4pt}
\textbf{VQA Circuit Parameter Encoding}.  
Prior machine learning–based approaches~\cite{fosel2021quantum,furrutterQuantumCircuitSynthesis2024} represent quantum circuits as three-dimensional tensors: the first dimension indexes qubits, the second specifies gate placement along the circuit depth, and the third denotes the gate type. Due to page limits, we omit details and refer readers to~\cite{fosel2021quantum,furrutterQuantumCircuitSynthesis2024}.  
We adopt the same representation but omit the third dimension, since the circuit ansatz is fixed for each task. This also enables a more compact depth dimension, as illustrated in Fig.~\ref{fig:design}(a). For instance, the two-qubit gate on $q_0$ need not appear at a later depth; all two-qubit gate parameters can be grouped together as long as the ansatz topology is known. This encoding provides a concise yet complete representation of circuit topology.

\textbf{Conditioning Text Encoding}.  
To guide parameter initialization across different VQAs, we use task-specific text prompts (Fig.~\ref{fig:design}(a)) that encode the defining features of each task. For example, in a variational quantum eigensolver (VQE) task, the prompt ``Hamiltonian = $0.5 \cdot IIZZ + ZXXZ$'' specifies the problem Hamiltonian, which uniquely characterizes the instance. The complete set of prompts for the five tasks studied in this work is given in Section~\ref{sec:exp} and summarized in Table~\ref{tab:dataset}. To convert these prompts into tensor representations suitable for conditioning a DDPM, we employ a pretrained CLIP encoder~\cite{radfordLearningTransferableVisual2021,ilharco_gabriel_openclip}, which aligns natural language with latent representations and enables text prompts to serve as effective conditioning inputs.

\vspace{-0.15in}
\subsection{\design~Architecture}
\vspace{-4pt}
At a high level, \design~is formulated as a DDPM that progressively corrupts VQA parameter tensors in the training set with varying levels of Gaussian noise. The noise prediction network is trained to estimate this noise so that subtracting it from the corrupted input reconstructs the original clean tensors. For the prediction network, we adopt a U-Net backbone~\cite{ronnebergerUNetConvolutionalNetworks2015}. The U-Net employs an encoder--decoder structure with skip connections, which simultaneously preserves fine-grained spatial information and captures hierarchical features. To tailor this architecture for VQA parameter initialization, we extend the original convolution-only design with several enhancements. 
\underline{First}, sinusoidal positional encoding (PE) layers are introduced to embed the diffusion time step $t$, the noisy input tensor $\mathbf{x}_{t}$, and the conditioning text embedding $\mathbf{c}$, thereby incorporating temporal and contextual information into the generative process. 
\underline{Second}, residual convolutional blocks~\cite{chenCrossViTCrossAttentionMultiScale2021} are integrated to improve feature extraction and training stability by facilitating gradient flow. 
\underline{Third}, transformer blocks~\cite{vaswaniAttentionAllYou2023} are embedded within the architecture to capture long-range dependencies and enhance representational capacity beyond the locality of convolutional operations.
The complete arrangement of these modules, including their depth and connectivity, is shown in Fig.~\ref{fig:design}(b).

\begin{table*}[!t]
    \footnotesize
    \caption{Dataset Overview}
    \vspace{-0.1in}
    \label{tab:dataset}
    \centering
    \setlength{\tabcolsep}{2pt}
    \begin{tabular}{|c|c|c|c|c|c|}
        \toprule
        \textbf{Application} & \textbf{Definition} & \textbf{Representative Prompt} & \textbf{\# Parameters} & \textbf{\# Qubits} & \textbf{\# Instances} \\\toprule
        \multicolumn{1}{|l|}{1D Heisenberg XYZ (1D\_XYZ)} & Equation~\eqref{eq:heisenberg_xyz} & \multicolumn{1}{|l|}{``$(J_1, J_2, J_3) = (2, 1, 0.5)$''} & 8 & 4 & 2000 \\\hline
        \multicolumn{1}{|l|}{1D Fermi--Hubbard Model (1D\_FH)} & Equation~\eqref{eq:1d_fh} & \multicolumn{1}{|l|}{``$(t, U) = (0.5, 1)$''} & 8 & 4 & 1000 \\\hline
        \multicolumn{1}{|l|}{2D Transverse-Field Ising Model (2D\_TFI)} & Equation~\eqref{eq:2d_tfi} & \multicolumn{1}{|l|}{``$(j, \mu) = (0.2, 3)$''} & 16 & 8 & 1000 \\\hline
      %  \ce{H2} & cf.~\cite{Utkarsh2023Chemistry} & ``Hydrogen molecule with bond length $0.74$ \AA'' & 24 & 4 & 150\\\hline
        \multicolumn{1}{|l|}{Quantum Pulse Synthesis (Q\_Pulse)} & cf.~\cite{liangNAPAIntermediatelevelVariational2024,wangQuantumNASNoiseAdaptiveSearch2022} & \multicolumn{1}{|l|}{``$h_0 = 0.3\,ZI + 0.2\,IZ;\; h_1 = 0.4\,XI;\; U_t = e^{-iHt}$''} & 24 & 2 & 8285 \\\hline
        \multicolumn{1}{|l|}{Random VQE} & cf.~\cite{liQASMBenchLowLevelQuantum2023} & \multicolumn{1}{|l|}{``Hamiltonian = $0.5 \cdot IIZZ + ZXXZ$''} & 48 & 4 & 2800 \\\bottomrule
    \end{tabular}
    \vspace{-0.22in}
\end{table*}

\vspace{-14pt}
\subsection{Training and Inference of \design}
\vspace{-4pt}

The training of~\design~follows the standard Markov chain formulation of DDPMs, comprising a forward process that progressively corrupts data with Gaussian noise and a reverse process that reconstructs clean samples from noisy inputs.

\textbf{Forward Process}.  
The forward (noising) process incrementally perturbs data with Gaussian noise according to a variance schedule \(\{\beta_t\}_{t=1}^T\):
{\setlength{\abovedisplayskip}{0pt}%
 \setlength{\belowdisplayskip}{0pt}%
 \begin{equation}\label{eq: ddpm_forward}
\begin{split}
      q(\mathbf{x}_{1:T}|\mathbf{x}_{0}) &=\prod_{t=1}^{T} q(\mathbf{x}_{t}|\mathbf{x}_{t-1}),\\
    q(\mathbf{x}_{t}|\mathbf{x}_{t-1})&=\mathcal{N}(\mathbf{x}_{t};\sqrt{1-\beta_t}\,\mathbf{x}_{t-1},\beta_t\mathbf{I}).  
\end{split}
\end{equation}}
With \(\alpha_t\coloneq1-\beta_t\) and \(\bar{\alpha}_t\coloneq\prod_{s=1}^t\alpha_s\), the marginal distribution admits a closed form:
{\setlength{\abovedisplayskip}{2pt}%
 \setlength{\belowdisplayskip}{2pt}%
 \begin{equation}\label{eq: ddpm_forward_closed}
    q(\mathbf{x}_t|\mathbf{x}_0)=\mathcal{N}(\mathbf{x}_t;\sqrt{\bar{\alpha}_t}\,\mathbf{x}_0,(1-\bar{\alpha}_t)\mathbf{I}),
\end{equation}}
which enables direct sampling at any timestep.  
\(\mathbf{x}_0\) denotes the optimized VQA parameters. As illustrated in Fig.~\ref{fig:design}(c), Gaussian noise is added batch-wise to parameter tensors, while prompt embeddings remain fixed. At each timestep \(t\), the tensor \(\mathbf{x}_{t}\) is thus obtained by further corrupting \(\mathbf{x}_{t-1}\).

Before feeding inputs into the model at timestep \(t\), we apply classifier-free guidance (CFG)~\cite{hoClassifierFreeDiffusionGuidance2022} to the prompt embeddings. Specifically, the text embedding \(\mathbf{c}\) is randomly mixed with a null embedding with probability \(p_{\text{guidance}}\). This balances the trade-off between fidelity (matching the intended VQA description) and diversity (encouraging broader exploration of ansatz parameters)~\cite{hoClassifierFreeDiffusionGuidance2022,dhariwalDiffusionModelsBeat2021}.

\textbf{Reverse Denoising Process}.  
The reverse process is parameterized and learned. Starting from pure Gaussian noise \(\mathbf{x}_T\sim\mathcal{N}(\mathbf{0},\mathbf{I})\), the generative model is expressed as
{\setlength{\abovedisplayskip}{2pt}
 \setlength{\belowdisplayskip}{2pt}
 \begin{equation}\label{eq: ddpm_reverse}    p_\theta(\mathbf{x}_{0:T})=p_\theta(\mathbf{x}_T)\prod_{t=1}^{T}p_\theta(\mathbf{x}_{t-1}|\mathbf{x}_t),
\end{equation}}
with Gaussian transitions of the form
{\setlength{\abovedisplayskip}{2pt}
 \setlength{\belowdisplayskip}{2pt}
 \begin{equation}
    p_\theta(\mathbf{x}_{t-1}|\mathbf{x}_t)=\mathcal{N}(\mathbf{x}_{t-1};\mu_\theta(\mathbf{x}_t,t),\Sigma_\theta(\mathbf{x}_t,t)).
\end{equation}}
Here, \(\theta\) denotes the parameters of the denoising network (the U-Net backbone), which predicts the conditional mean \(\mu_\theta\) and variance \(\Sigma_\theta\). At each timestep, the U-Net receives as input the noisy tensor \(\mathbf{x}_t\), the timestep \(t\), and the CFG-modified text embedding \(\mathbf{c}\), and predicts the noise tensor \(\hat{\epsilon}_\theta\). Sampling proceeds iteratively from \(t=T\) to \(0\), generating a denoised parameter tensor \(\mathbf{x}_0\).

\textbf{Training Loss}.  
\design~parameters \(\theta\) are optimized by minimizing the variational bound on the negative log-likelihood, which reduces to a denoising score-matching objective~\cite{hoDenoisingDiffusionProbabilistic2020}:
{\setlength{\abovedisplayskip}{2pt}
 \setlength{\belowdisplayskip}{2pt}
 \begin{equation}\label{eq:ddpm_loss}
    L_{\mathrm{MSE}}=\mathbb{E}_{\mathbf{x}_0,t,\epsilon}\Big[\big\|\epsilon-\epsilon_\theta(\sqrt{\bar{\alpha}_t}\mathbf{x}_0+\sqrt{1-\bar{\alpha}_t}\epsilon,t)\big\|^2\Big],
\end{equation}}
where
{\setlength{\abovedisplayskip}{2pt}
 \setlength{\belowdisplayskip}{2pt}
 \begin{equation}
    \epsilon=\frac{\mathbf{x}_t-\sqrt{\bar{\alpha}_t}\mathbf{x}_0}{\sqrt{1-\bar{\alpha}_t}}, 
    \epsilon_\theta=\frac{\sqrt{1-\bar{\alpha}_t}}{\beta_t}\big(\mathbf{x}_t-\sqrt{\alpha_t}\,\mu_\theta(\mathbf{x}_t,t)\big).
\end{equation}}
In this formulation, \(\epsilon_\theta(\cdot,t)\) is the neural network’s prediction of the noise added at timestep \(t\). Once the prediction error between \(\epsilon_\theta\) and the true noise \(\epsilon\) is minimized across all timesteps, the training of \design~is complete.

\textbf{Inference}.  
At inference time, given an unseen VQA task, the objective is to generate a parameter tensor \(\mathbf{x}_{0}\) that provides a high-quality initialization for the corresponding VQA ansatz. As depicted in Fig.~\ref{fig:design}(d), the textual description of the task is first mapped to an embedding \(\mathbf{c}\) using the OpenCLIP encoder. Classifier-free guidance (CFG) is applied to \(\mathbf{c}\) to reinforce semantic alignment with the task description while retaining generative diversity. Initialization begins with a Gaussian noise tensor \(\mathbf{x}_{T} \sim \mathcal{N}(\mathbf{0}, \mathbf{I})\), which is progressively denoised by the trained U-Net from timestep \(T\) to \(0\), conditioned on the embedding \(\mathbf{c}\). The resulting tensor \(\mathbf{x}_{0}\) is then decoded via the inverse of the circuit encoding scheme in Fig.~\ref{fig:design}(a) to recover the VQA ansatz parameters.

\begin{figure*}[t]
\centering
\setlength{\tabcolsep}{1pt} % no extra col spacing
\renewcommand{\arraystretch}{0.0}
\begin{tabular}{@{}c cccc@{}}
% -------- Row 1 --------
\raisebox{0.45\height}{%
  \rotatebox{90}{\parbox{2.2cm}{\centering\textbf{\small Init Loss Dist}}}%
}\hspace{-2pt} &
\includegraphics[width=0.32\linewidth]{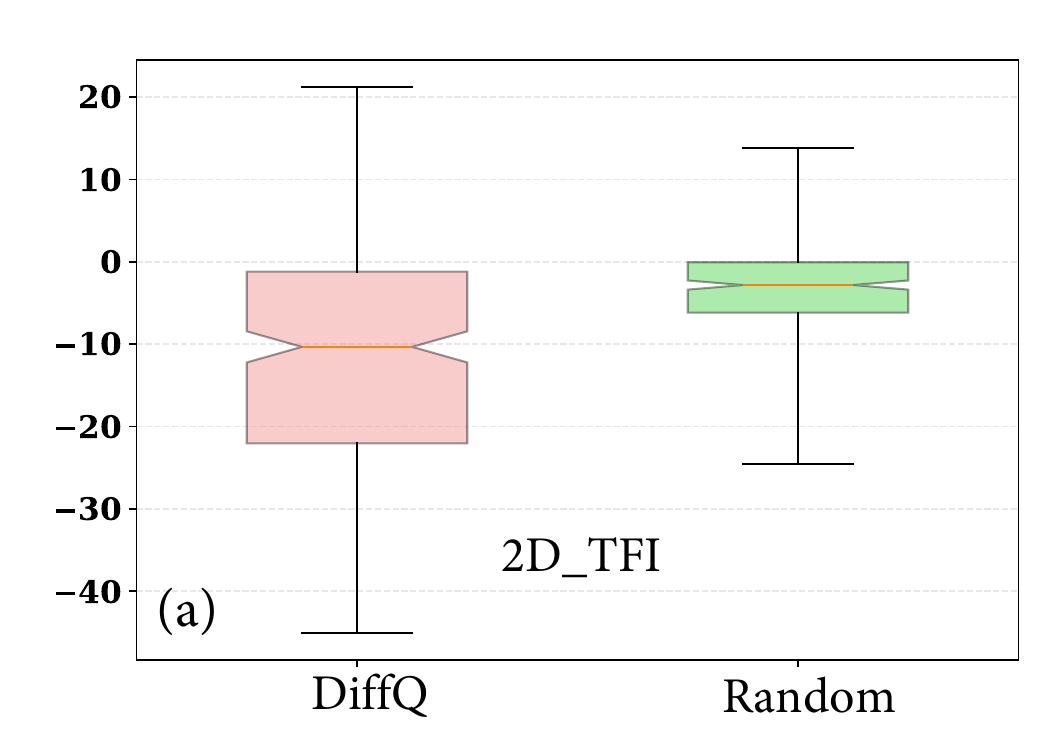} &
\includegraphics[width=0.32\linewidth]{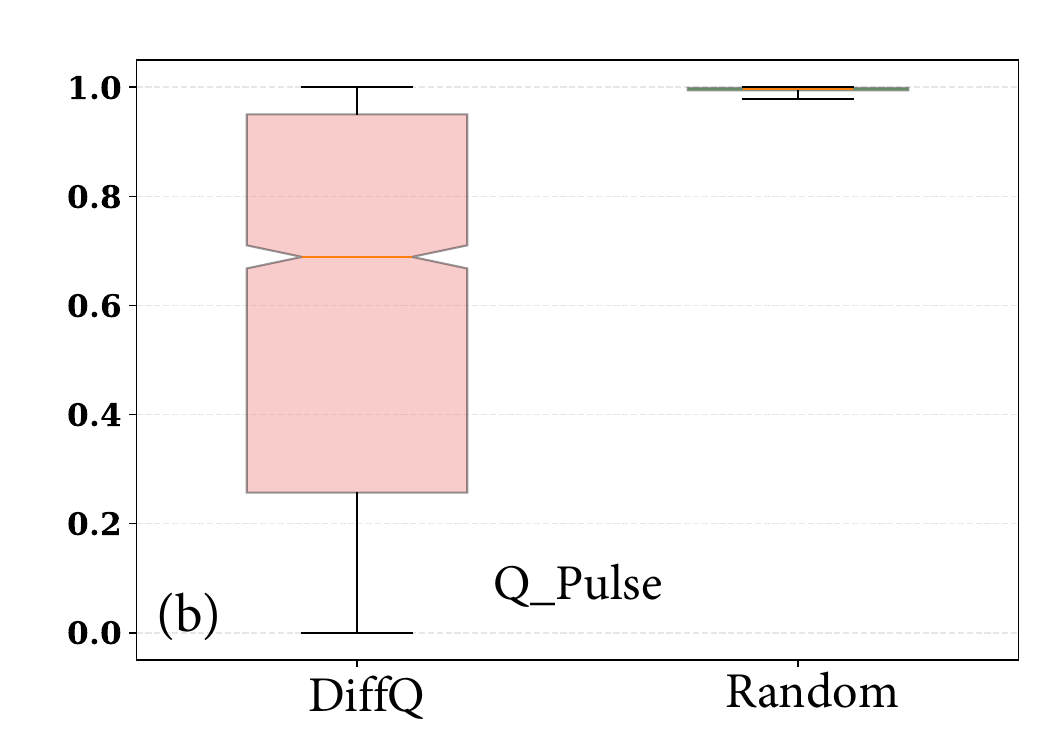} &
\includegraphics[width=0.32\linewidth]{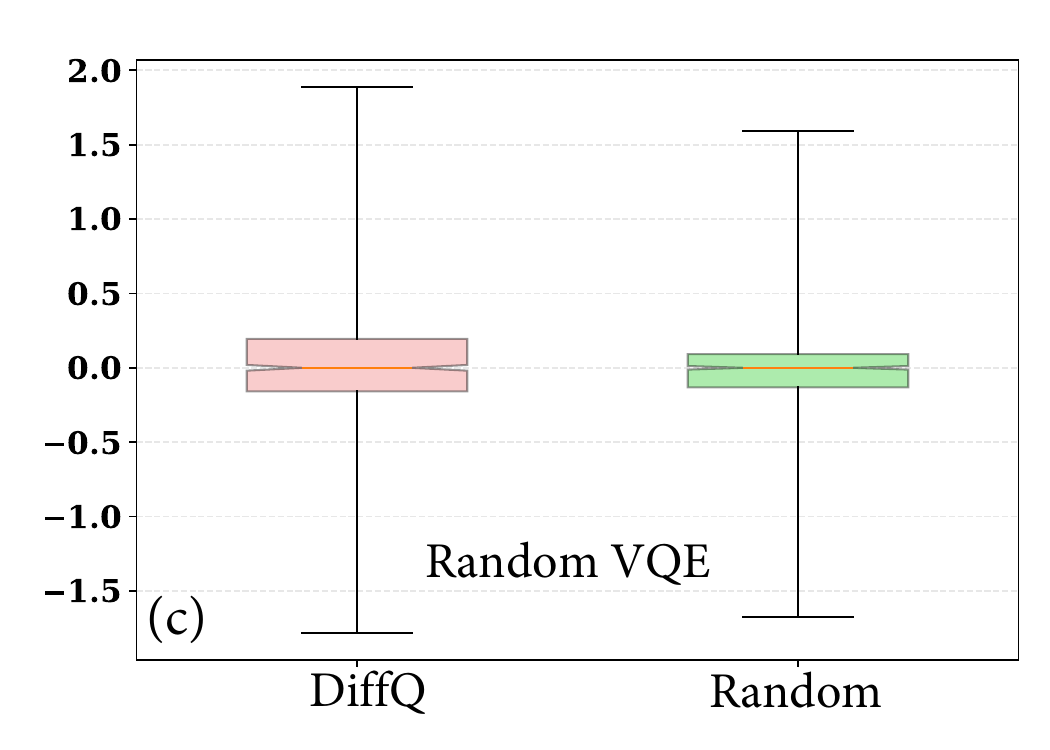} \\
% -------- Row 2 --------
\raisebox{0.45\height}{%
  \rotatebox{90}{\parbox{2.2cm}{\centering\textbf{\small Loss Curves}}}%
}\hspace{-2pt} &
\includegraphics[width=0.32\linewidth]{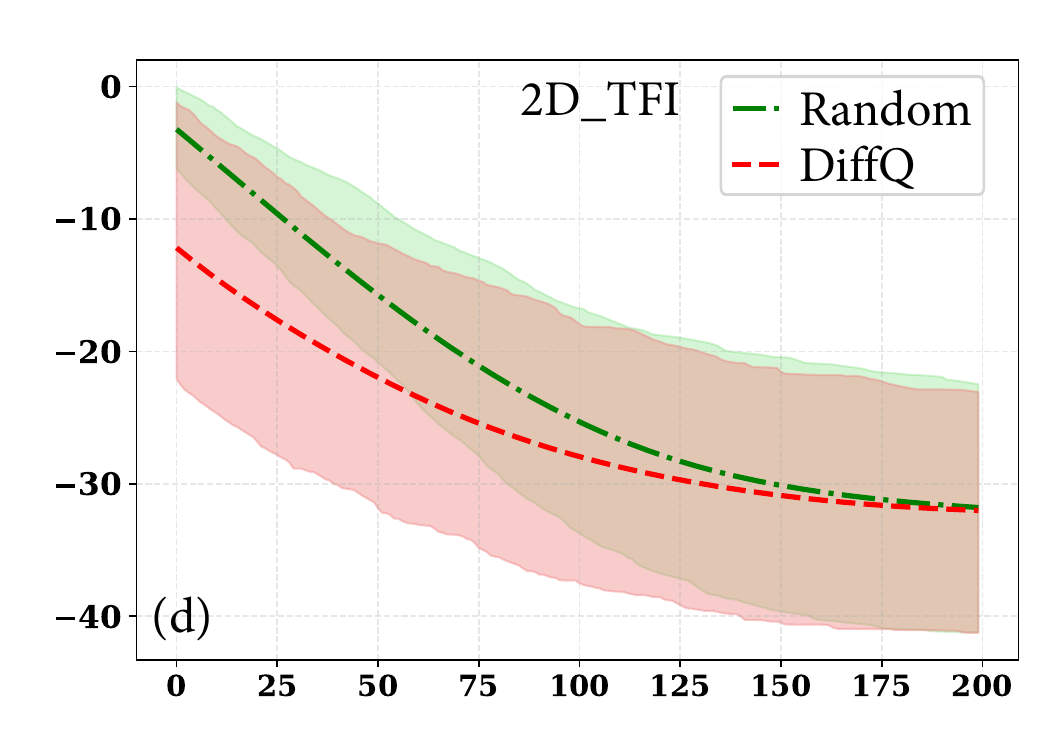} &
\includegraphics[width=0.32\linewidth]{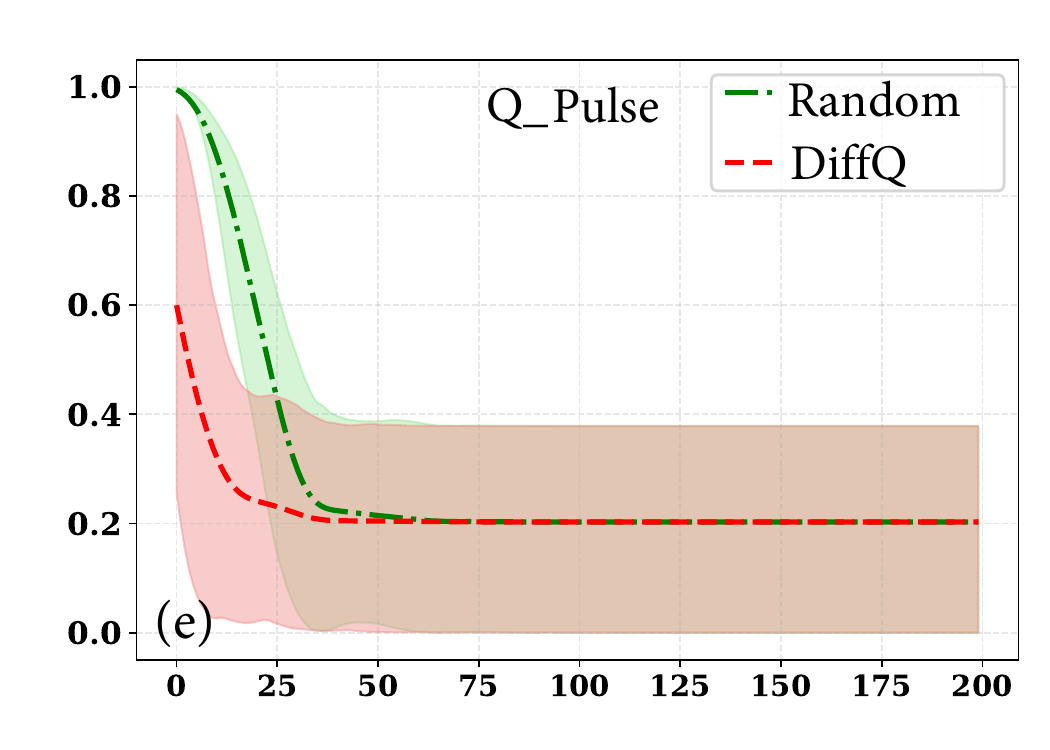} &
\includegraphics[width=0.32\linewidth]{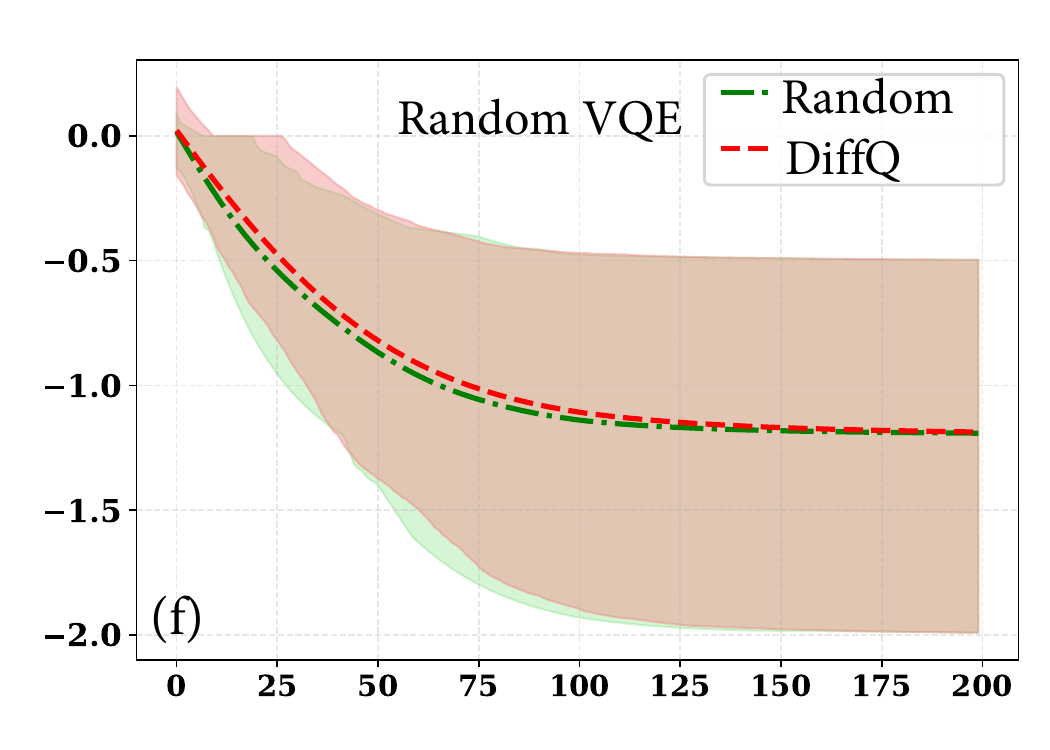} \\
\end{tabular}
\vspace{-12pt}
\caption{Initial loss distributions and training loss trajectories for three distinct tasks.}
\label{fig:init_loss_loss_curve}
\vspace{-18pt}
\end{figure*}

\vspace{8pt}
\section{Dataset Construction}
\label{sec:exp}
\vspace{-12pt}

We construct comprehensive VQA datasets spanning three domains with five representative tasks, each paired with optimized ansatz parameters and a corresponding text prompt template capturing key task features. Each dataset comprises thousands of instances, in contrast to prior work~\cite{mesman2024nn,liangGraphLearningParameter2024,mengParameterGenerationQuantum2024,leeQMAMLQuantumModelAgnostic2025}, which typically provide only hundreds of samples.  
The first domain is \textit{quantum many-body physics}, represented by three tasks: the Heisenberg XYZ model (1D\_XYZ) with Hamiltonian in Eq.~\eqref{eq:heisenberg_xyz}; the Fermi--Hubbard model (1D\_FH) with Hamiltonian in Eq.~\eqref{eq:1d_fh}; and the Transverse--Ising model (2D\_TFI) with Hamiltonian in Eq.~\eqref{eq:2d_tfi}.  
Beyond many-body physics, we introduce two additional domains: \textit{quantum pulse synthesis} (Q\_Pulse), as described in~\cite{liangNAPAIntermediatelevelVariational2024,wangQuantumNASNoiseAdaptiveSearch2022,lowOptimalArbitrarilyAccurate2014,brownArbitrarilyAccurateComposite2004,merrillProgressCompensatingPulse2012}, and \textit{general random variational quantum eigensolvers} (VQE), detailed in~\cite{liQASMBenchLowLevelQuantum2023}.  
Instances are generated by varying problem parameters, with circuits optimized using standard techniques in \texttt{PennyLane} and \texttt{TorchQuantum}~\cite{wangQuantumNASNoiseAdaptiveSearch2022}. Dataset statistics and key features are summarized in Table~\ref{tab:dataset}.

\vspace{-10pt}
{
\setlength{\abovedisplayskip}{-1pt}
\setlength{\belowdisplayskip}{-1pt}
\setlength{\abovedisplayshortskip}{-1pt}
\setlength{\belowdisplayshortskip}{-1pt}
\begin{align}
\label{eq:heisenberg_xyz}
    \hat{H} &= \sum_{i=0}^3(J_1 X_i X_{i+1} + J_2 Y_i Y_{i+1} + J_3 Z_i Z_{i+1}), \\
\label{eq:1d_fh}
    \hat{H} &= -t\sum_{i=0}^{n-1} (\hat{c}^\dagger_i \hat{c}_{i+1} + \hat{c}_{i+1}\hat{c}) + U\sum_{i=0}^{n-1}\hat{n}_i \hat{n}_{i+1},\\
\label{eq:2d_tfi}
    \hat{H} &= -j \sum_{i, j}Z_i Z_j - \mu \sum_{i} Z_i.
\end{align}
}

\vspace{-0.12in}
\section{Experimental Setup and Results}
\label{sec:results}
\vspace{-0.1in}

\subsection{Experimental Setup}
\vspace{-6pt}

We conduct experiments to evaluate the performance of \design~on the constructed datasets. The evaluation metrics follow prior work~\cite{mesman2024nn,liangGraphLearningParameter2024,mengParameterGenerationQuantum2024,leeQMAMLQuantumModelAgnostic2025}, namely the average initial loss and the average number of convergence steps in VQA optimization. Since existing methods~\cite{mesman2024nn,liangGraphLearningParameter2024,mengParameterGenerationQuantum2024,leeQMAMLQuantumModelAgnostic2025} are limited to a single application domain (e.g., many-body physics or quantum pulse synthesis), we adopt random initialization as a general baseline for comparison.  
For each application, all instances are randomly split into training and testing sets with a ratio of $7:3$. We train \design~on each application independently and evaluate its performance on the corresponding test set. Unless otherwise specified, the diffusion step is set to $T=100$, the guidance scale to $g=10$, and the noise schedule to linear, ranging from $\beta_0=0.0001$ to $\beta_T=0.02$. Each model is trained for $500$ epochs with a maximum learning rate of $5 \times 10^{-5}$ on a single NVIDIA RTX 3090 GPU.

\subsection{Experimental Results}
\vspace{-6pt}

%%%%%%%%%%%%%%%%%%%%%%%%%%%%%%%%%%%%%%%%%%%%%%%%%%%%%%%
\begin{table}[t]
\footnotesize
\vspace{4pt}
\caption{Comparison of~\design~with random initialization. $\Delta$ denotes the relative difference, where positive values indicate improvement and negative values indicate degradation.}
\label{tab:eval_metrics}
\vspace{-8pt}
\centering
\setlength{\tabcolsep}{6pt}
\begin{tabular}{|c| c| c| c|}
\toprule
\textbf{Applications} 
& \textbf{Schemes} 
& \textbf{Initial Loss} 
& \textbf{Convergence Steps} \\\toprule

\multirow{3}{*}{\textbf{1D\_XYZ}}
& Random    &0.06   & 266.93  \\\cline{2-4}
& \design  &-0.16  &251.89  \\\cline{2-4}
& \textbf{$\Delta$}  
&\color{red}{$0.22\downarrow$}  
&\color{red}{$5.6\%\downarrow$} \\\hline
\multirow{3}{*}{\textbf{1D\_FH}}
& Random     & 1.01  &218.46 \\\cline{2-4}
& \design  &0.47  &213.21 \\\cline{2-4}
& \textbf{$\Delta$}  
&\color{red}{$0.54\downarrow$}  
&\color{red}{$2.4\%\downarrow$} \\\hline
\multirow{3}{*}{\textbf{2D\_TFI}}
& Random    &-3.23  & 235.93 \\\cline{2-4}
& \design  &-12.18   &180.83 \\\cline{2-4}
& \textbf{$\Delta$}  
&\color{red}{$8.95\downarrow$}  
&\color{red}{$23.4\%\downarrow$} \\\hline
\multirow{3}{*}{\textbf{Q\_Pulse}}
& Random    &0.99   & 93.21  \\\cline{2-4}
& \design  &0.60 &81.98  \\\cline{2-4}
& \textbf{$\Delta$}  
&\color{red}{$0.39\downarrow$}  
&\color{red}{$12\%\downarrow$} \\\hline
%
%\multirow{3}{*}{\textbf{\ce{H2} Molecule}}
%& Random     &-0.22  & 215.91\\\cline{2-4}
%& \design  &-0.36  &600 \\\cline{2-4}
%& \textbf{$\Delta$}  
%&\color{red}{$0.14\downarrow$}  
%& \color{blue}{$177.9\%\uparrow$}\\\hline
%
\multirow{3}{*}{\textbf{Random VQE}}
& Random     &0.02  &237.82 \\\cline{2-4}
& \design  &0.02  &250.26 \\\cline{2-4}
& \textbf{$\Delta$}  
& \color{red}{$0$}   
&\color{red}{$5.2\%\uparrow$} \\\bottomrule
\end{tabular}
\vspace{-0.15in}
\end{table}
%%%%%%%%%%%%%%%%%%%%%%%%%%%%%%%%%%%%%%%%%%%%%%%%%%%%%%%

\textbf{Overall Performance Comparison}.  
Table~\ref{tab:eval_metrics} summarizes the performance of~\design~relative to random initialization, evaluated in terms of initial loss and convergence steps across five datasets. All values are averaged over the corresponding test sets, with relative differences from random initialization highlighted in red.  
Overall,~\design~consistently outperforms the baseline by achieving lower initial loss (with reductions of up to 8.95) and faster convergence (requiring up to 23.4\% fewer optimization steps). These improvements are particularly significant for NISQ devices, where faster convergence reduces exposure to quantum noise and lowers computational cost.  
For the Random VQE tasks, however,~\design~performs comparably to or slightly worse than random initialization. This limitation arises because such tasks lack meaningful physical correlations, thereby reducing the utility of learned parameter priors. These observations indicate that~\design~is most effective in settings where domain-specific structures are present, underscoring its relevance for realistic quantum applications that demand both scalability and precision.  

\textbf{Initial Loss and Convergence Analysis}.  
Figure~\ref{fig:init_loss_loss_curve} further compares~\design~with random initialization across three representative applications: 2D\_TFI, Q\_Pulse, and Random VQE. Figures~\ref{fig:init_loss_loss_curve}(a–c) show the distributions of initial losses across all test instances, while Figures~\ref{fig:init_loss_loss_curve}(d–f) depict the corresponding convergence trajectories. In the convergence plots, the shaded regions denote the range of losses across all instances, and the dashed curves represent pointwise averages.  

For the 2D\_TFI and Q\_Pulse tasks,~\design~achieves lower initial loss and faster convergence than random initialization. In the Q\_Pulse case, random initialization causes degenerate behavior with losses collapsing near one, whereas~\design~avoids this issue and accelerates convergence.  
By contrast, in the Random VQE task,~\design~shows no clear advantage, as both initial loss and convergence closely match random initialization. This reflects the task’s simplicity (a single Pauli term), where learned priors are unnecessary. Overall, these results demonstrate that~\design~is highly effective in structured, domain-informed applications but offers limited benefits for trivial or unstructured problems.

\vspace{-0.1in}
\section{Conclusion}
\vspace{-0.1in}
We present~\design, a unified framework for generating high-quality ansatz parameters in variational quantum algorithms. Based on a denoising diffusion probabilistic model,~\design~iteratively transforms random noise into task-specific ansatz parameters conditioned on textual descriptions of the target problem. Experiments show that~\design~consistently surpasses random initialization, achieving lower initial loss and faster convergence across diverse quantum applications.

% \section*{Acknowledgment}
% \TODO{add grants and acknowledgments.}

%\newpage

\clearpage

\let\oldbibliography\thebibliography
\renewcommand{\thebibliography}[1]{\oldbibliography{#1}
\setlength{\itemsep}{-3pt}}

\bibliographystyle{IEEEbib.bst}
\bibliography{reference}

\end{document}